\documentclass[conference]{IEEEtran}
\usepackage[utf8]{inputenc}
\usepackage[T1]{fontenc}
\usepackage[numbers,sort&compress]{natbib}
\usepackage{booktabs}
\usepackage{graphicx}
\usepackage{xcolor}
\usepackage{amssymb}
\usepackage{url}
\usepackage[hidelinks]{hyperref}
\usepackage{cleveref}
\usepackage[framemethod=tikz]{mdframed}
\crefname{section}{Section}{Sections}
\Crefname{section}{Section}{Sections}
\crefname{table}{Table}{Tables}
\crefname{figure}{Figure}{Figures}

\newcommand{\retro}{\texttt{/retro}}
\mdfdefinestyle{rqanswer}{leftline=true,rightline=false,topline=false,bottomline=false,linewidth=2pt,innerleftmargin=8pt,innerrightmargin=0pt,innertopmargin=2pt,innerbottommargin=2pt,leftmargin=0pt,rightmargin=0pt,skipabove=\topsep,skipbelow=\topsep}
\newcommand{\rqanswer}[2]{\begin{mdframed}[style=rqanswer]\textbf{Answer to #1:} #2\end{mdframed}}

\begin{document}

\title{How Do Coding-Agent Harnesses Self-Evolve?\\An Empirical Study of Retrospective Adaptation}

\author{\IEEEauthorblockN{Christoph Treude}
\IEEEauthorblockA{Singapore Management University\\
Singapore\\
ctreude@smu.edu.sg}
\and
\IEEEauthorblockN{Sebastian Baltes}
\IEEEauthorblockA{Heidelberg University\\
Heidelberg, Germany\\
sebastian.baltes@uni-heidelberg.de}}

\maketitle

\begin{abstract}
Coding agents can adapt their harnesses, the software, configuration, and instructions used to run them, based on experience from earlier tasks, but whether these adaptations improve their behavior on later tasks remains unclear. We study \retro{}, a skill that asks an agent to review a completed session and propose harness changes, and ask what changes it proposes and whether implemented changes reduce repeated work on later tasks. We ran Claude Code on 30 SWE-bench Verified issues, qualitatively coded the 125 proposals from \retro{}, implemented 24, and evaluated 12 harness changes on later issues from the same repositories in fresh sessions with and without each change. Most proposals referred to an event in the session, but 18 of 43 lint-related proposals recommended a practice regardless of what had happened. Among the seven evaluations in which sessions without the change encountered the targeted problem, two changes eliminated the targeted action, one reduced its occurrence, four showed no reduction, and some changes added work on unrelated issues. Harness self-evolution requires more than generating and implementing changes: it needs ways to establish whether a change helps on later tasks and to revise or remove changes that do not.
\end{abstract}


\section{Introduction}
\label{sec:intro}

Coding agents such as Claude Code run inside a harness: the software, configuration, and instructions used to run the agent, including context files such as \texttt{CLAUDE.md} that the agent reads at the start of each session. A session that identifies how to run a repository's tests or which tests already fail can retain this information in the harness for later sessions on the same repository. Retaining changes from completed sessions for use on later tasks, without retraining the model, is harness self-evolution \cite{zhou2026secasurvey}.
Whether a retained change helps is a separate question: a change can describe one task rather than the repository, can stay outside a later session's context, or can be read without changing what the agent does.

The \retro{} skill is one mechanism for producing such changes. It instructs a coding agent to conduct a retrospective: a review of a completed session that generates numbered proposals for harness changes requiring human approval \cite{pocock2026skills}.
In our study, Claude Code completed the task sessions, generated the retrospectives, and implemented approved proposals. Two large language models (LLMs) coded the proposals, and we selected which proposals to implement, chose the later issues, and labeled the outcomes, so the process we study is self-evolution with human selection and evaluation rather than an autonomous loop. We ask two research questions. RQ1: What harness changes does \retro{} propose, and to what extent are they based on events in the session? RQ2: Under what conditions do implemented adaptations prevent targeted actions on later issues from the same repository?

An implemented harness change is an adaptation. For each adaptation, we operationalize repeated work as a later session performing a specific action that the adaptation was intended to prevent, such as running tests with \texttt{pytest} when it is not installed, or rerunning tests on unmodified code to check whether an unrelated failure was already present. This is a behavioral measure, not a judgment that the action was unnecessary or that preventing it improved the task outcome. We transferred individual adaptations, each proposed after one session, to other issues in the same repository. By recording whether later sessions encountered the targeted problem, consulted the adaptation, and repeated the targeted action, we examine when a retained adaptation changes what the agent does.

\section{Background and related work}
\label{sec:related}

Previous studies examined how developers configured coding agents and whether the configuration helped. \citeauthor{galster2026harness} found context files in 91\% of the 2,853 engineered GitHub projects with configuration artifacts for coding tools \cite{galster2026harness}.
\citeauthor{lulla2026agentsmd} ran Codex on 124 pull requests from 10 repositories with and without each repository's \texttt{AGENTS.md} file, and median wall-clock time fell from 99 to 70 seconds with the file \cite{lulla2026agentsmd}.
On 58 SWE-bench Verified tasks, \citeauthor{zhang2026guardrails} observed the same pass rate for random community rule files and curated rules \cite{zhang2026guardrails}.
These studies supplied instructions written or generated before the sessions.

Other work retained experience from earlier attempts. Agentic Harness Engineering uses session traces to propose harness edits, and its authors checked the predicted effects of those edits in later benchmark rounds \cite{lin2026ahe}.
\citeauthor{wu2026modularrsi} traced harness changes into later attempts at the same task, but those attempts used several changes together. They warned that an update based on one session trace could address reasoning specific to a task rather than a recurring harness problem \cite{wu2026modularrsi}.
\citeauthor{lin2026updating} compared task success with original and updated harnesses and examined whether agents loaded skills into context and followed their instructions \cite{lin2026updating}.
In a BenchTrace experiment that transferred experience between different subtasks of a text adventure, all six methods achieved less task progress than the baseline without memory updates \cite{huang2026benchtrace}.
Previous work has examined whether updated harnesses help and whether agents load and follow their instructions. We complement these studies by evaluating individual adaptations, one at a time, on other issues from the same repository.

\section{Study design}
\label{sec:design}

We first ran 30 coding-agent sessions and collected 125 proposals from \retro{}. We then coded the proposals, implemented 24 of them, and evaluated 11 resulting adaptations and one author-written comparison adaptation on later issues from the same repositories. All task sessions ran non-interactively using Claude Code 2.1.295 with Opus~5.5 and the same task prompt template, using \retro{} from source commit \texttt{b0618bc} \cite{pocock2026skills}.
The artifact contains the prompts, codebook, labeling instructions, complete data, and the procedural details omitted below.

As an example of what \retro{} produces, consider a SymPy session that attempted to run tests using \texttt{pytest}, absent from the container, before finding the repository's \texttt{bin/test} runner. The second of its four proposals, headed ``No pointer to the test runner,'' observed that the repository had no context file and proposed a two-line \texttt{CLAUDE.md} naming \texttt{bin/test}, stating that \texttt{pytest} was not installed, and listing the CI checks to run before finishing. Claude Code implemented the file unchanged. Its intended effect was that later sessions in the repository would run \texttt{bin/test} first.

\textbf{Tasks and sessions.} The exploratory sample comprised 10 issues, two each from Django, SymPy, Matplotlib, Astropy, and Sphinx, sampled at random from the benchmark's ``15 min--1 hour'' category of estimated developer effort \cite{swebenchverified2024}.
The second sample of 20 issues from Astropy, pytest, Sphinx, and SymPy was not random: we chose repositories in which the exploratory sessions had encountered environment problems, such as pre-existing test failures, and issues with another issue opened later in the same repository. The proposal counts therefore describe these 30 sessions, not arbitrary coding tasks. Each issue ran in a fresh Claude Code session in a Docker container with the issue's repository snapshot and installed dependencies. The prompt supplied the issue description without additional hints and asked Claude Code to modify the source code and verify its changes using existing tests. After Claude Code finished, we resumed the same session with \retro{} and saved the output unchanged, regardless of whether the patch passed the hidden benchmark tests. For the exploratory sample, our experiment script committed the patch before invoking \retro{}, and we excluded three remarks about that commit from the analysis.

\textbf{Proposal coding.} Each numbered entry in the output from \retro{} counted as one proposal. All 125 proposals were coded qualitatively using a fixed codebook. The \emph{trigger} described the relationship between a proposal's rationale and the events recorded in the session trace: \emph{experience-driven} marked proposals based on an event observed in that trace, \emph{prior-driven} marked proposals justified without relying on session events, such as adding a lint hook solely because the repository lacked one, and \emph{weak} marked an indirect connection to the session. The \emph{mechanism} described the type of change that a proposal recommended, using the six codes listed in \Cref{tab:proposals}. Two coders without access to each other's labels produced the initial coding: Claude Code agents with model \texttt{claude-fable-5-1} and default settings, working from the full session traces, and GPT with model \texttt{gpt-6.1-sol} and high reasoning effort, working from trace excerpts. Both coders used the same codebook, so their agreement reflects consistency across these two inputs, not human inter-rater reliability. They assigned the same trigger code to 109 of 125 proposals (Cohen's $\kappa$ = 0.67) and the same mechanism code to 109 ($\kappa$ = 0.84). The first author resolved the 91 disagreements, counted per proposal and code, and we report the resolved labels. The artifact documents the coders' inputs and the review of the exploratory sample.

\textbf{Eligibility and implementation.} A proposal was eligible for implementation and evaluation on another issue if it was coded \emph{experience-driven}, proposed a persistent harness change in permitted files and settings, including context files, hooks, and lint or continuous integration (CI) configuration, was safe to apply, did not merely fix the original issue, and admitted a written criterion for identifying repeated work in later sessions. The initial coding by Claude identified 33 eligible proposals, 14 from the exploratory sample and 19 from the second sample. The first author reviewed the exploratory-sample eligibility labels before implementation, and reviewed the second-sample labels only after implementation, where the two coders disagreed. From the second sample, we selected nine known-failures files and one test-command instruction, the two types for which other issues reproduced the targeted problem. Further SymPy proposals of these types were omitted as already evaluated in that repository, and for the remaining proposals we found no other issue reproducing the targeted problem. The original Claude Code sessions were resumed in their original containers to implement these 10 and all 14 exploratory-sample proposals. Eleven of the 24 implementations produced adaptations that we could evaluate on another issue. The other 13 changed no harness files, changed only the environment or files outside the permitted locations, or had no other issue reproducing the targeted problem. For comparison with these experience-driven adaptations, we also evaluated one pre-commit configuration from a proposal coded \emph{prior-driven}. The configuration was written by the authors after Claude Code's permission system blocked its implementation, giving 12 evaluated adaptations. The resolution of disagreements changed the eligibility of 16 proposals, none of them among the 12 evaluated adaptations.

\textbf{Transfer.} We call each evaluation of one adaptation on a task selected to reproduce the problem targeted by that adaptation a \emph{transfer event}. For each event, we chose another SWE-bench Verified issue from the same repository and verified in a fresh Docker container that its environment reproduced the problem, for example that the listed tests failed on unmodified code. This targeted selection tests whether an adaptation changes behavior where its problem can arise, not how often it would help on arbitrary later issues. Five sessions ran under the original harness and five under the adapted harness, each in a fresh container with the same task prompt and, for the adapted harness, the changed files copied without modification. We evaluated 12 adaptations in 13 transfer events, testing one adaptation on two tasks. Three adaptations were also evaluated on control tasks, unrelated issues from the same repositories, to check whether they added verification work. The two Sphinx adaptations shared one control issue.

For each session, we recorded \emph{recurrence}: whether Claude Code performed the action that the adaptation was intended to prevent, such as trying \texttt{pytest} first when it was not installed. For pre-existing failures, recurrence meant a \emph{baseline comparison}: comparing test results from edited and unmodified code to determine whether an observed failure already existed. We excluded comparisons limited to the issue's target tests or tests added by Claude Code, and comparisons against a stored failure list alone. A session that encountered a failure absent from its known-failures file and compared against unmodified code performed the action the file was meant to prevent, although the comparison was a reasonable verification step. A problem present in a task's environment was also not necessarily encountered by a session: a stale leap-second data file only failed tests that the session chose to run, so a session that ran only the issue's target tests never encountered it. We therefore distinguish encountering a problem from coverage. A session \emph{encountered} the problem when its trace showed the class of problem targeted by the adaptation, such as any pre-existing test failure. The adaptation \emph{covered} the problem when it addressed the particular instances encountered, such as the specific failing tests. We classified a transfer event as \emph{informative} when at least three of its five original-harness sessions encountered the problem, regardless of coverage. We marked an adaptation as \emph{loaded} when it entered Claude Code's context and as \emph{consulted} when Claude Code read a referenced file or ran a recommended command. Consulting a file did not establish that Claude Code acted on its contents. For the nine events concerning pre-existing failures and their control tasks, two LLM labelers and then the first author labeled all 120 session traces for recurrence and encountered failures without labels identifying the harness. The reported labels are the first author's, in agreement with the LLM labels on recurrence in 119 of 120 sessions ($\kappa$ = 0.98). One adapted session that ended on an authentication error was excluded, and we report counts of valid sessions, with task success and elapsed time as secondary outcomes.
\section{Results}
\label{sec:results}

\textbf{Sessions and proposals (RQ1).} Each retrospective produced two to six proposals, 125 in total (\Cref{tab:proposals}). Of the 125 proposals, 101 were coded \emph{experience-driven}, 19 \emph{prior-driven}, and 5 \emph{weak}. Seventeen of the 19 \emph{prior-driven} proposals were executable checks. Only nine of the 38 executable checks could report failure in the task environment as written. Of the other 29, 28 needed missing tools or were commit hooks that never fired because the task sessions made no commits, and one offered a \texttt{CLAUDE.md} pointer, a change that cannot fail, as the alternative to installing \texttt{pytest}.

\begin{table}[t]
\caption{Mechanism codes of the 125 proposals from \retro{}: number of proposals, number of the 30 sessions with at least one such proposal, and one example proposal with its trigger code. \emph{None} marks a change to how Claude Code performs a task with no change to its harness.}
\label{tab:proposals}
\centering
\scriptsize
\setlength{\tabcolsep}{3pt}
\renewcommand{\arraystretch}{0.9}
\begin{tabular}{@{}lccp{3.6cm}@{}}
\toprule
Mechanism & Prop. & Sess. & Example (trigger) \\
\midrule
Executable check & 38 & 29 & Pre-commit hook running \texttt{flake8} on staged files (SymPy; \emph{prior-driven}) \\
Navigation pointer & 24 & 22 & \texttt{CLAUDE.md} line naming \texttt{bin/test} as the test runner (SymPy; \emph{experience-driven}) \\
Instruction & 21 & 16 & Coding-standards rule that a bug fix includes a test failing before the fix (Astropy; \emph{experience-driven}) \\
Information access & 20 & 18 & Known-failures file listing tests failing under the installed docutils (Sphinx; \emph{experience-driven}) \\
Tool or configuration & 12 & 10 & Script printing one line per failing test instead of tracebacks (SymPy; \emph{experience-driven}) \\
None & 10 & 10 & Use the Edit tool instead of a Python replace script (SymPy; \emph{experience-driven}) \\
\bottomrule
\end{tabular}
\end{table}

Claude Code produced 43 lint-related proposals across 29 of 30 sessions, counting SymPy's \texttt{bin/test quality} command as linting. We coded 23 of them as \emph{experience-driven}, 18 as \emph{prior-driven}, and 2 as \emph{weak}, and lint accounted for 18 of the 19 \emph{prior-driven} proposals overall. The skill's instructions treat a repository with neither a pre-commit hook nor a CI job running its checks as ``a finding'' \cite{pocock2026skills}.
Retrospectives therefore do more than translate recorded events into harness changes. They also recommend practices the skill considers good regardless of whether their absence caused a problem in the session. Environment problems recurred across sessions. The traces of 5 of 10 exploratory sessions and 19 of 20 second-sample sessions contained a failed shell command, a repeated shell command, or more than one test runner, almost always because of pre-existing test failures from incompatible docutils or Python versions or outdated leap-second data, or missing lint tools. Claude Code proposed known-failures files in 15 of 30 sessions to avoid repeating baseline comparisons for pre-existing test failures. Test-runner instructions and known-failures files supplied all 11 evaluated adaptations from \emph{experience-driven} proposals.

\rqanswer{RQ1}{Executable checks were the most common change (38 of 125), and 101 proposals were based on events in the session, but 18 of the 19 \emph{prior-driven} proposals recommended lint checks regardless of what had happened.}

\textbf{Transfer (RQ2).} Of the 13 transfer events, seven were \emph{informative} (\Cref{tab:transfer}). In five events, the issue environment reproduced the targeted problem, but no original-harness session encountered it in four events and only one of five did in the fifth. The thirteenth event was the comparison adaptation, the pre-commit configuration we wrote from a \emph{prior-driven} proposal: the file was present in all five adapted sessions, but no session read it or ran pre-commit. Claude Code loaded the transferred context-file instructions in every adapted session and usually consulted the referenced files. \Cref{tab:transfer} groups the informative events by whether the test command or the listed failures covered the instances encountered.

Two adaptations eliminated repeated work in all adapted sessions. The SymPy \texttt{CLAUDE.md} line from the example in \Cref{sec:design} prevented the initial attempt to run the unavailable \texttt{pytest} in every adapted session, reducing recurrence from 5 of 5 sessions to 0 of 5. In Astropy, a known-failures file listed three failing tests, one involving stale leap-second data and two involving a NumPy deprecation. All ten sessions encountered one of these failures, and adapted sessions read the file instead of running baseline comparisons, reducing recurrence from 3 of 5 sessions to 0 of 5.

In four transfer events, recurrence did not decrease. In each of them, sessions encountered failures absent from the transferred known-failures files. Each of two Sphinx files listed two failing tests and described unlisted failures as new. All adapted sessions read the files, encountered failures in other tests, and ran baseline comparisons. A SymPy known-failures file and comparison script covered two packages, but sessions tested other packages, and a pytest known-failures file listed three failures in one test file, but sessions encountered failures in other files. An Astropy adaptation combined a plugin to repair the leap-second cache with a script that compared test failures against a stored list. Recurrence fell from 5 of 5 sessions to 2 of 5, but no session encountered the leap-second failure. The sessions encountered NumPy failures outside the script's stored failure list, which it reported as new.

The control tasks showed that an adaptation can also add work. One Sphinx known-failures file told Claude Code to treat any failure absent from the file as caused by its own edits. On an unrelated control issue, four of five adapted sessions ran baseline comparisons, compared with none of five original-harness sessions. An incomplete failure list was thus followed by the verification it was meant to prevent. On the other two control tasks, 1 of 5 and 0 of 5 adapted sessions did so, against none under the original harness.

\begin{table}[t]
\caption{Informative transfer events: valid sessions with recurrence under each harness (control tasks: baseline comparisons) and adapted sessions that consulted the adaptation. Events are grouped by whether the adaptation covered the particular instances that the sessions encountered.}
\label{tab:transfer}
\centering
\scriptsize
\setlength{\tabcolsep}{2.5pt}
\renewcommand{\arraystretch}{0.9}
\begin{tabular}{llccc}
\toprule
Repository & Adaptation & Original & Adapted & Consulted \\
\midrule
\multicolumn{5}{l}{\textit{Adaptation covered encountered instances}} \\
SymPy & \texttt{CLAUDE.md} line naming \texttt{bin/test} & 5/5 & 0/5 & 5/5 \\
Astropy & known-failures file: 3 tests & 3/5 & 0/5 & 5/5 \\
\quad control task & same file, unrelated issue & 0/5 & 0/5 & 0/5 \\
\midrule
\multicolumn{5}{l}{\textit{Adaptation did not cover encountered instances}} \\
Sphinx & known-failures file + \texttt{AGENTS.md} & 5/5 & 5/5 & 5/5 \\
\quad control task & same file, unrelated issue & 0/5 & 1/5 & 5/5 \\
Sphinx & known-failures file + \texttt{CLAUDE.md} & 5/5 & 5/5 & 5/5 \\
\quad control task & same file, unrelated issue & 0/5 & 4/5 & 5/5 \\
SymPy & 2-package known-failures file + script & 4/5 & 4/4 & 4/4 \\
pytest & known-failures file: 3 tests & 5/5 & 5/5 & 5/5 \\
Astropy & cache repair plugin + script & 5/5 & 2/5 & 5/5 \\
\bottomrule
\end{tabular}
\end{table}

Task success differed by at most one session between harnesses in every event, providing no consistent evidence of improved task success. Lower recurrence did not consistently mean shorter sessions, in particular when an adaptation added instructions. The SymPy \texttt{CLAUDE.md} file removed the failed \texttt{pytest} call but bundled a second line telling Claude Code to run the CI lint checks before finishing. Every adapted session ran them, and median elapsed time rose from 88 to 156 seconds. It also rose in the SymPy and pytest events with uncovered failures and on the Sphinx control task, where it went from 82 to 245 seconds. It fell by 5 to 28 seconds in the other four informative events.

\rqanswer{RQ2}{The two adaptations that covered the instances later sessions encountered eliminated the targeted action; of the five that did not, one reduced it and four left it unchanged, one of which added work on an unrelated issue.}

\section{Threats to validity}
\label{sec:threats}

The selection of evaluation issues that reproduced the targeted problem may overestimate how often adaptations would help on unselected tasks. Claude Code implemented proposals in the sessions and containers used to generate them, so the changed files could include information absent from the proposal text. The traces could reveal Claude Code reading transferred files, so the first author's labeling was not fully blind. With five sessions per harness and seven informative events, the counts in \Cref{tab:transfer} describe these cases and do not establish that covering the encountered problem explains the differences between adaptations, which differ in several properties at once and were grouped after labeling.

\section{Conclusion}
\label{sec:conclusion}

We collected empirical evidence on how coding-agent harnesses self-evolve through retrospectives. For RQ1, most proposals referred to an event in the session, but lint proposals often recommended a practice regardless of what had happened. For RQ2, implementing an adaptation did not guarantee that it reduced repeated work.

These observations suggest that retrospective harness adaptation must account for the scope of retained experience. A repository's test command can remain useful across issues, whereas a list of the specific tests that failed in one environment may not cover those encountered later. An evaluation of an adaptation therefore needs to establish whether the targeted problem arose, whether the agent consulted the adaptation, and whether its actions changed. Retained changes also need continued evaluation, because outdated or incomplete instructions can remain ineffective or add work.

Our evidence covers one agent, one version of \retro{}, and environment problems. Future work should evaluate other agents, adaptations for code navigation and coding conventions, and the components of bundled adaptations one at a time, on later issues not selected for the targeted problem.

\section*{Data availability}
An artifact with all scripts, prompts, session traces, codebook, labels, and harness changes is available at \url{https://doi.org/10.5281/zenodo.23653700}.

\bibliographystyle{IEEEtranN}

\begin{thebibliography}{10}
\providecommand{\natexlab}[1]{#1}
\providecommand{\url}[1]{#1}
\csname url@samestyle\endcsname
\providecommand{\newblock}{\relax}
\providecommand{\bibinfo}[2]{#2}
\providecommand{\BIBentrySTDinterwordspacing}{\spaceskip=0pt\relax}
\providecommand{\BIBentryALTinterwordstretchfactor}{4}
\providecommand{\BIBentryALTinterwordspacing}{\spaceskip=\fontdimen2\font plus
\BIBentryALTinterwordstretchfactor\fontdimen3\font minus
  \fontdimen4\font\relax}
\providecommand{\BIBforeignlanguage}[2]{{%
\expandafter\ifx\csname l@#1\endcsname\relax
\typeout{** WARNING: IEEEtranN.bst: No hyphenation pattern has been}%
\typeout{** loaded for the language `#1'. Using the pattern for}%
\typeout{** the default language instead.}%
\else
\language=\csname l@#1\endcsname
\fi
#2}}
\providecommand{\BIBdecl}{\relax}
\BIBdecl

\bibitem[Zhou et~al.(2026)Zhou, Hu, Luo, Shang, Fang, Chen, Xiao, and
  Zhang]{zhou2026secasurvey}
H.~Zhou, H.~Hu, T.~Luo, Y.~Shang, C.~Fang, Z.~Chen, L.~Xiao, and Q.~Zhang,
  ``Self-evolving coding agents,'' arXiv:2608.03392, 2026.

\bibitem[poc(2026)]{pocock2026skills}
``skills: Agent skills for engineering, productivity, and writing,''
  \url{https://github.com/mattpocock/skills}, commit b0618bc, 2026, accessed
  2026-10-08.

\bibitem[Galster et~al.(2026)Galster, Mohsenimofidi, Lulla, Abubakar, Treude,
  and Baltes]{galster2026harness}
M.~Galster, S.~Mohsenimofidi, J.~L. Lulla, M.~A. Abubakar, C.~Treude, and
  S.~Baltes, ``Harness engineering for agentic {AI} coding tools: An
  exploratory study,'' arXiv:2602.14690, 2026.

\bibitem[Lulla et~al.(2026)Lulla, Mohsenimofidi, Galster, Zhang, Baltes, and
  Treude]{lulla2026agentsmd}
J.~L. Lulla, S.~Mohsenimofidi, M.~Galster, J.~M. Zhang, S.~Baltes, and
  C.~Treude, ``On the impact of {AGENTS.md} files on the efficiency of {AI}
  coding agents,'' arXiv:2601.20404, 2026, presented at the Journal Ahead
  Workshop (JAWs) at ICSE 2026.

\bibitem[Zhang et~al.(2026)Zhang, Wang, Cui, Qiu, Li, Zhu, and
  He]{zhang2026guardrails}
X.~Zhang, G.~Wang, Y.~Cui, W.~Qiu, Z.~Li, B.~Zhu, and P.~He, ``Guardrails beat
  guidance: A large-scale study of rules, skills, and persistent configuration
  for coding agents,'' arXiv:2604.11088, 2026.

\bibitem[Lin et~al.(2026{\natexlab{a}})Lin, Liu, Pan, Lin, Dou, Xi, Huang, Yan,
  Han, Gui, and Jiang]{lin2026ahe}
J.~Lin, S.~Liu, C.~Pan, L.~Lin, S.~Dou, Z.~Xi, X.~Huang, H.~Yan, Z.~Han,
  T.~Gui, and Y.-G. Jiang, ``Agentic harness engineering: Observability-driven
  automatic evolution of coding-agent harnesses,'' arXiv:2604.25850, 2026.

\bibitem[Wu et~al.(2026)Wu, Ren, Li, Li, Yang, Ma, Zhang, Gu, Yang,
  Batista-Navarro, Zhang, Liu, Zhou, Dai, and Lin]{wu2026modularrsi}
S.~Wu, J.~Ren, Y.~Li, H.-S. Li, C.~Yang, X.~Ma, Y.~Zhang, W.~Gu, J.~Yang,
  R.~Batista-Navarro, C.~Zhang, X.~Liu, M.~Zhou, B.~Dai, and C.~Lin,
  ``{ModularRSI}: Modular and generalizable recursive harness
  self-improvement,'' arXiv:2609.14857, 2026.

\bibitem[Lin et~al.(2026{\natexlab{b}})Lin, Wu, Wang, Shi, Sang, He, Liu, Wei,
  Wu, Zhang, Wang, Zhang, Dumoulin, Xie, Zhou, Wang, and Lu]{lin2026updating}
M.~Lin, J.~Wu, Z.~Wang, Z.~Shi, Y.~Sang, B.~He, Z.~Liu, T.~Wei, Z.~Wu,
  Z.~Zhang, D.~Wang, X.~Zhang, B.~Dumoulin, C.~Xie, Y.~Zhou, S.~Wang, and
  H.~Lu, ``Harness updating is not harness benefit: Disentangling evolution
  capabilities in self-evolving {LLM} agents,'' arXiv:2605.30621, 2026.

\bibitem[Huang et~al.(2026)Huang, Cheng, Jiang, Yu, and
  Aizawa]{huang2026benchtrace}
J.~Huang, F.~Cheng, J.~Jiang, Z.~Yu, and A.~Aizawa, ``{BenchTrace}: A benchmark
  for testing reflection ability and controlled evolution in {LLM} agents,''
  arXiv:2605.29225, 2026.

\bibitem[{OpenAI}(2024)]{swebenchverified2024}
{OpenAI}, ``Introducing {SWE-bench} {Verified},''
  \url{https://openai.com/index/introducing-swe-bench-verified/}, 2024,
  accessed 2026-10-10.

\end{thebibliography}

\end{document}